\documentclass[a4paper,11pt]{article}\usepackage{jcappub}\def\PRDstyle#1{}\def\JCAPstyle#1{#1}\let\Abstract\abstract

\usepackage{graphicx}
\usepackage[utf8]{inputenc}
\usepackage[T1]{fontenc}
\usepackage{cmap}

\def\imo{i}

\def\im#1{Im(#1)}

\def\D{{\cal D}}
\def\P{\Psi}
\def\PD{\bar{\Psi}}
\def\gc{\gamma}
\def\ge{\hat{\gamma}}
\def\K{{\cal K}}

\begin{document}
\title{Wormholes without exotic matter: quasinormal modes, echoes and shadows}
\JCAPstyle{
\author[\star]{M. S. Churilova,}\emailAdd{wwrttye@gmail.com}
\author[\star\dagger]{R. A. Konoplya,}\emailAdd{roman.konoplya@gmail.com}
\author[\star]{Z. Stuchlík,}\emailAdd{zdenek.stuchlik@physics.slu.cz}
\author[\star\ddagger]{A. Zhidenko}\emailAdd{olexandr.zhydenko@ufabc.edu.br}
\affiliation[\star]{Research Centre for Theoretical Physics and Astrophysics,\\ Institute of Physics, Silesian University in Opava,\\ Bezručovo náměstí 13, CZ-74601 Opava, Czech Republic}
\affiliation[\dagger]{Peoples Friendship University of Russia (RUDN University),\\ 6 Miklukho-Maklaya Street, Moscow 117198, Russian Federation}
\affiliation[\ddagger]{Centro de Matemática, Computação e Cognição (CMCC),\\ Universidade Federal do ABC (UFABC),\\ Rua Abolição, CEP: 09210-180, Santo André, SP, Brazil}
\arxivnumber{2107.05977}
}
\PRDstyle{
\author{M. S. Churilova}\email{wwrttye@gmail.com}
\affiliation{Research Centre for Theoretical Physics and Astrophysics, Institute of Physics, Silesian University in Opava, Bezručovo náměstí 13, CZ-74601 Opava, Czech Republic}
\author{R. A. Konoplya}\email{roman.konoplya@gmail.com}
\affiliation{Research Centre for Theoretical Physics and Astrophysics, Institute of Physics, Silesian University in Opava, Bezručovo náměstí 13, CZ-74601 Opava, Czech Republic}
\affiliation{Peoples Friendship University of Russia (RUDN University), 6 Miklukho-Maklaya Street, Moscow 117198, Russian Federation}
\author{Z. Stuchlík}\email{zdenek.stuchlik@physics.slu.cz}
\affiliation{Research Centre for Theoretical Physics and Astrophysics, Institute of Physics, Silesian University in Opava, Bezručovo náměstí 13, CZ-74601 Opava, Czech Republic}
\author{A. Zhidenko}\email{olexandr.zhydenko@ufabc.edu.br}
\affiliation{Research Centre for Theoretical Physics and Astrophysics, Institute of Physics, Silesian University in Opava, Bezručovo náměstí 13, CZ-74601 Opava, Czech Republic}
\affiliation{Centro de Matemática, Computação e Cognição (CMCC), Universidade Federal do ABC (UFABC),\\ Rua Abolição, CEP: 09210-180, Santo André, SP, Brazil}
}

\Abstract{
An analytical solution representing traversable asymptotically flat and symmetric wormholes was obtained without adding exotic matter in two different theories independently: in the Einstein-Maxwell-Dirac theory and in the second Randall-Sundrum brane-world model. Further, a smooth normalizable asymmetric wormhole solution has been recently obtained numerically in the Einstein-Maxwell-Dirac theory. Using the time-domain integration method we study quasinormal ringing of all these wormholes with emphasis to the regime of mimicking the near extremal Reissner-Nordström black holes, which is characterised by echoes. In addition, we calculate radius of shadows cast by these wormholes.
}
\maketitle

\section{Introduction}

Wormholes are structures which connect disparate points of spacetime or different universes \cite{Visser:1995cc}. Existence of traversable wormholes requires gravitational repulsion, preventing the throat from shrinking. This repulsion could be supported by some exotic matter. Usually wormholes without adding exotic matter are constructed at the price of modifications of the underlaying gravitational theory \cite{McFadden:2004ni,Bronnikov:2002rn,Bronnikov:2003gx,Richarte:2007zz,Kanti:2011jz,Kanti:2011yv,Bronnikov:2019sbx,Maldacena:2020sxe}. However, recently a class of asymptotically flat, symmetric relatively the throat and traversable wormhole solutions was obtained in General Relativity with added Maxwell and two Dirac fields with usual coupling between them~\cite{Blazquez-Salcedo:2020czn}. There, two kinds of solutions were obtained: the analytical solution supported by massless, neutral fermions and the numerical solutions (when the fermion is massive). The first solution corresponds to the fermion wave function which does not decay at infinity and is, therefore, non-normalizable. The second, numerical, solution is normalizable, but has a number of other peculiar properties, such as the requirement of posing a thin shell on the wormhole throat with nonphysical properties: a) fermion particles and anti-particles must meet at the throat without annihilation, b) the matter fields are not smooth on the shell. The smooth asymptotically flat wormhole solutions, which are free of all the above problems, have been recently found in \cite{Konoplya:2021hsm}. These solutions describe wormholes which are asymmetric relatively the throat.

Having in mind that the analytical solution obtained in \cite{Blazquez-Salcedo:2020czn} is supported by non-normalizable massless Dirac field which approaches a constant at infinity, we, nevertheless, can interpret it as a reasonable approximation to normalizable case, when the mass of the fermion is very small (neutrino). The numerical solutions found in \cite{Blazquez-Salcedo:2020czn,Konoplya:2021hsm} indeed show that once the fermion mass is small the geometry changes insignificantly. After all, the same analytical solution, up to the redefinition of constants, was obtained in the Randall-Sundrum brane-world model (RS2) \cite{Bronnikov:2002rn} by K.~Bronnikov and S.~Kim. Thus, the analytical wormhole solution will be considered here as a separate case appearing in two different theories. The numerical asymmetric wormhole solution, supported by the normalizable and smooth matter fields~\cite{Konoplya:2021hsm} will be another case of viable wormholes constructed without adding exotic matter.

Here we will concentrate on two main effects in the vicinity of wormholes: proper oscillation frequencies given by quasinormal modes and shadows cast by wormholes observable in the electromagnetic spectrum. The term shadow for wormhole means certainly the shape of the photon sphere as seen by a distant observer, rather than the black spot cast by a black hole. Both quasinormal modes of various wormholes and their shadows were studied in quite a few papers and our analysis will be complementary to that performed in earlier works~\cite{Konoplya:2005et,Konoplya:2010kv,Bronnikov:2012ch,Konoplya:2016hmd,Konoplya:2018ala,Churilova:2019qph,Damour:2007ap,Jusufi:2021lei,Peng:2021osd,Javed:2020mjb,Wielgus:2020uqz,Shaikh:2018oul,Amir:2018pcu,Gyulchev:2018fmd,Ohgami:2015nra,Jusufi:2020mmy,Roy:2019yrr,Blazquez-Salcedo:2018ipc,Aneesh:2018hlp,Volkel:2018hwb,Bueno:2017hyj,Nandi:2016uzg,Cardoso:2016rao,Churilova:2019cyt}. This concerns, first of all, the calculation of shadow radius for the Kim-Bronnikov-2 wormhole~\cite{Bronnikov:2002rn} done in~\cite{Bronnikov:2021liv}. Quasinormal modes for the Kim-Bronnikov-2 wormhole were computed for the $\ell=1$ multipole of a test scalar field in \cite{Bronnikov:2019sbx} with a numerical mistake and here we will correct it and generalize computations for various multipole numbers and fields (scalar, Dirac, electromagnetic).
In this aspect an important feature of the wormhole geometries under consideration is that in the particular limit of parameter values they approach the black-hole state, so that near the transition point, the wormhole mimic the black-hole behavior~\cite{Damour:2007ap,Cardoso:2016rao,Churilova:2019cyt,Bronnikov:2021liv}. Thus, we are aimed at the thorough study of the three phenomena: quasinormal modes, echoes and shadows of wormholes constructed without exotic matter.

Our paper is organized as follows: In Sec.~\ref{sec:metrics} we give basic information on analytical and numerical wormhole solutions under consideration. Sec.~\ref{sec:waveequations} is devoted to master wave equations for test scalar, electromagnetic and Dirac fields in the wormhole's vicinity. Sec.~\ref{sec:QNMs} discusses the quasinormal modes and echoes from wormholes. Sec.~\ref{sec:shadows} is devoted to calculation of the shadow radius. Finally, in the Conclusion we summarize the obtained results and mention some open questions.

\section{Metrics}\label{sec:metrics}

\subsection{Einstein-Dirac-Maxwell (EDM) model}

In \cite{Blazquez-Salcedo:2020czn} a model with two gauged {relativistic} fermions is considered, the spin of which is taken to be opposite in order to satisfy spherical symmetry.
We will use the units $G=c=\hbar=1$. Then the action of the corresponding Einstein-Dirac-Maxwell (EDM) model reads
\begin{eqnarray}
\label{action}
S = \frac{1}{4\pi} \int \mathrm d^4 x \sqrt{-g} \, \left[ \frac{1}{4}R + \mathcal L_1 + \mathcal L_2 -\frac{1}{4}F^2 \right] \, ,
\end{eqnarray}
where $R$ is the Ricci scalar of the metric
$g_{\mu\nu}$,
$$F_{\mu\nu}=\partial_{\mu}A_\nu-\partial_{\nu}A_\mu$$
is the field strength tensor of the U(1) field $A_\mu$, and
\begin{eqnarray}
\label{Lagrangian_Dirac}
\mathcal L_\epsilon = \frac{\imo}{2}\PD_\epsilon\gc^{\mu}\D_{\mu}\P_\epsilon-\frac{\imo}{2}(\PD_\epsilon\gc^{\mu}\D_{\mu}\P_\epsilon)^*-\imo\mu\PD_\epsilon\P_\epsilon,
\end{eqnarray}
where $\gc^\mu$ are the curved space gamma matrices\footnote{We use the notation and conventions in Ref.~\cite{Dolan:2015eua}.} and $\mu$ is the mass of both spinors $\Psi_{\boldsymbol{\epsilon}=1,2}$.
In addition, we have
\begin{equation}\label{Diracadjoint}
\D_{\mu}\equiv\partial_{\mu} - \Gamma_\mu - \imo qA_{\mu},
\end{equation}
where $q$ is the gauge coupling constant (charge) and $\Gamma_\mu$ are the spinor connection matrices,
\begin{equation}\label{spinorconnection}
\Gamma_\mu=-\frac{1}{4}\omega_{\mu\alpha\beta}\ge^{\alpha}\ge^{\beta},
\end{equation}
which are calculated by making use of the spin connection
\begin{equation}\label{spinconnection}
\omega_{\mu\alpha\beta}=(e_{\nu\alpha}\Gamma^{\nu}_{\mu\rho}-\partial_{\mu}e_{\rho\alpha})g^{\rho\lambda}e_{\lambda\beta}.
\end{equation}
Here $\Gamma^{\nu}_{\mu\rho}$ is the Levi-Civita connection, $e_{\mu\alpha}$ are the tetrads and $\ge^{\alpha}$ are the Dirac matrices,
$$e_{\mu\alpha}e_{\nu\beta}\eta^{\alpha\beta}=g_{\mu\nu},\qquad\ge^{\alpha}\ge^{\beta}+\ge^{\beta}\ge^{\alpha}=2\eta^{\alpha\beta}.$$

The resulting field equations are
\begin{eqnarray}\label{Dirac}
\left(\gc^{\nu}\D_{\nu}-\mu\right)\P_{\epsilon}&=&0,\\
\label{Maxwell}
\frac{\partial_{\nu}\sqrt{-g}F^{\mu\nu}}{2\sqrt{-g}}&=&q (\PD_1\gc^{\mu}\P_1+\PD_2\gc^{\mu}\P_2),\\
\label{Einstein}
G_\mu^\nu\equiv R_\mu^\nu - \frac{1}{2}\delta_\mu^\nu R&=&2F^{\nu\lambda}F_{\mu\lambda}-\frac{1}{2}\delta_\mu^\nu F^2+T_\mu^\nu,
\end{eqnarray}
where the stress-energy tensor for the Dirac field is
\begin{eqnarray}\label{DiracT}
T_{\mu\nu}&=&2\sum\limits_{\epsilon=1,2}\im{\PD_{\epsilon}\gc_{\mu}\D_{\nu}\P_{\epsilon}+\PD_{\epsilon}\gc_{\nu}\D_{\mu}\P_{\epsilon}}.
\end{eqnarray}

A spherically symmetric configuration is given by the following line element and four-potential:
\begin{eqnarray}\nonumber
&&ds^2=-N^2(r)dt^2 + \frac{dr^2}{B^2(r)} + r^2(d\theta^2 + \sin^2\theta d\varphi^2),\\
&&A_{\mu}dx^{\mu}=V(r)dt,
\end{eqnarray}
and the tetrads can be chosen as
\begin{equation}\label{vielbein}
e_{\mu\alpha}=\left(
                  \begin{array}{cccc}
                    -N(r) & 0 & 0 & 0 \\
                    0 & 1/B(r) & 0 & 0 \\
                    0 & 0 & r & 0 \\
                    0 & 0 & 0 & r\sin\theta \\
                  \end{array}
                \right).
\end{equation}

According to \cite{Herdeiro:2017fhv}, we use the following ansatz for the spinors:
\begin{eqnarray}\label{spinoransatz1}
\Psi_1 &=& e^{-\imo\omega t+\imo\varphi/2}\left(\begin{array}{r}
\phi(r)\cos\frac{\theta}{2} \Bigr.\\
\imo\kappa\phi^*(r)\sin\frac{\theta}{2} \Bigr.\\
-\imo\phi^*(r)\cos\frac{\theta}{2} \Bigr.\\
-\kappa\phi(r)\sin\frac{\theta}{2} \Bigr.
\end{array}\right),\\ \label{spinoransatz2}
\Psi_2 &=& e^{-\imo\omega t-\imo\varphi/2}\left(\begin{array}{r}
\imo\phi(r)\sin\frac{\theta}{2} \Bigr.\\
\kappa\phi^*(r)\cos\frac{\theta}{2} \Bigr.\\
\phi^*(r)\sin\frac{\theta}{2} \Bigr.\\
\imo\kappa\phi(r)\cos\frac{\theta}{2} \Bigr.
\end{array}\right),
\end{eqnarray}
with $\kappa=\pm1$ and
\begin{equation}
\phi(r)=e^{\imo\pi/4}F(r)-e^{-\imo\pi/4}G(r),
\end{equation}

The resulting EDM equations can be solved analytically in the $q=0$ limit, the spinor fields being massless ($\mu=0$), with $\omega=0$.
The solution has the following metric and U(1) potential
\begin{eqnarray}
\nonumber
N(r)&=&1-\frac{2Q_e^2}{Q_e^2+r_0^2}\frac{r_0}{r},\\
\label{analyticsln}
B(r)&=&\pm\sqrt{\left(1-\frac{Q_e^2}{r_0r}\right)\left(1-\frac{r_0}{r}\right)},\\\nonumber
V(r)&=&\frac{2Q_er_0}{Q_e^2+r_0^2}B(r),\\\nonumber
F(r)&=&\frac{c_0}{\sqrt{N(r)}}\left(\sqrt{1-\frac{Q_e^2}{r_0r}}-\kappa\sqrt{1-\frac{r_0}{r}}\right)^2,\\\nonumber
G(r)&=&\frac{\kappa r_0(Q_e^2+r_0^2)^{-1}}{32 c_0\sqrt{N(r)}}\left(\sqrt{1-\frac{Q_e^2}{r_0r}}+\kappa\sqrt{1-\frac{r_0}{r}}\right)^2,
\end{eqnarray}
with $c_0\neq 0$ being an arbitrary constant.
This configuration describes a (regular) traversable wormhole solution, with the throat's radius $r_0$ and the electric charge $|Q_e|<r_0$, while the ADM mass is
$$M\equiv\frac{2Q_e^2r_0}{Q_e^2+r_0^2}<|Q_e|.$$

The wormhole geometry is supported by the spinors contribution to the total energy-momentum tensor,
being regular everywhere.
As $Q_e\to r_0$,
the extremal Reissner-Nordström black hole is approached, while
$T_{\mu\nu}\to 0$.

\subsection{Asymmetric wormhole in Einstein-Dirac-Maxwell theory}

It was shown in \cite{Konoplya:2021hsm} that smooth traversable wormholes supported by the massive fermions are asymmetric. The corresponding numerical solutions can be parametrized by the values of $b_i$, where
$$b_i^2 = \lim_{r\to r_0}r\frac{dB^2}{dr}.$$
As $b_i$ approaches zero, the wormhole becomes more symmetric and the geometry on both sides of the throat approaches the extreme Reissner-Nordström black holes. In addition, for given value of $b_i$ there are different solutions, supported by distinctive fermion configurations, corresponding to different values of the parameters
$$f_i \equiv F(r_0)\sqrt{r_0}, \qquad g_i \equiv G(r_0)\sqrt{r_0}.$$
The wormholes' asymmetry is characterized by different asymptotic masses $M_{\pm}$, charges $Q_{\pm}$, and the redshift, such that the redshifted observer measures smaller asymptotic mass $M_-$ and charge $Q_-$ (in units of the wormhole's throat size). We present some relevant parameters of such asymmetric wormholes supported by the massive charged fermions ($qr_0=0.03$ and $\mu r_0=0.2$) in Table~\ref{tabl:shadowradius} and refer the reader to \cite{Konoplya:2021hsm} for more details.

Unlike the symmetric configurations proposed in \cite{Blazquez-Salcedo:2020czn}, the asymmetric wormhole spacetimes constructed in \cite{Konoplya:2021hsm} are the solutions of the Einstein-Dirac-Maxwell equations in the whole space. Since all the corresponding functions, describing the metric and matter fields, are smooth in this case, no specific membrane of matter at the throat or matching conditions are required to support such wormholes \cite{Konoplya:2021hsm} and, consequently, the criticism by \cite{Bolokhov:2021svp,Danielson:2021aor} towards \cite{Blazquez-Salcedo:2020czn} is limited by the symmetric case. At the same time, the metric given by eqs.~(\ref{analyticsln}) with non-normalizable spinors is smooth across the throat. This metric appears also as an exact solution in the brane-world model without matter fields and, naturally, without the matching problem across the throat. Thus within the brane-world approach this analytical solution is smooth and carries no disadvantages of the  Einstein-Dirac-Maxwell theory. Therefore, here we will study the analytical solution, having in mind its brane-world interpretation.

\begin{figure*}
\resizebox{\linewidth}{!}{\includegraphics{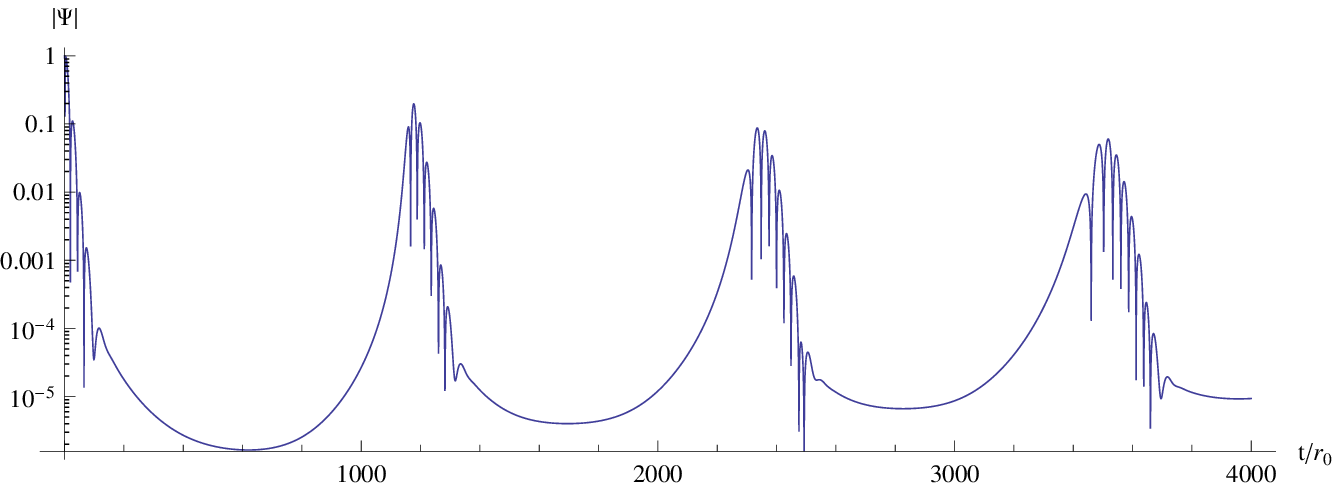}}
\resizebox{\linewidth}{!}{\includegraphics{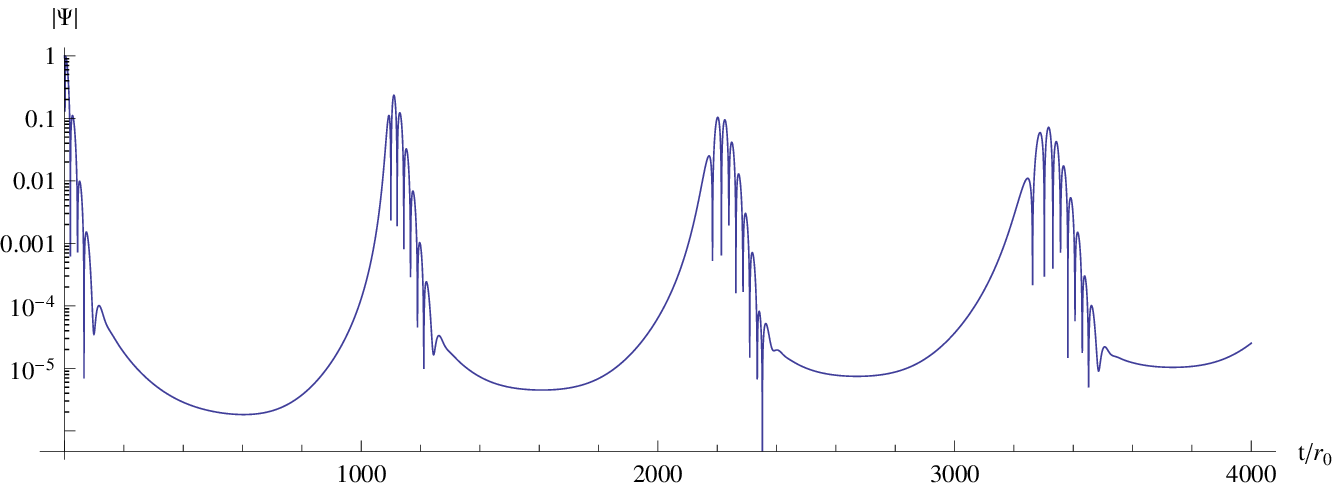}}
\caption{Time-domain profile for the $\ell=0$ scalar field perturbation in the background of the asymmetric wormhole ($b_i=0.14$, $f_i=-0.0193696$, $g_i=0.0153756$) supported by the massive charged fermions ($qr_0=0.03$ and $\mu r_0=0.2$). We present two cases: when the perturbation and observer location are on the ``$+$'' (blueshifted) side of the wormhole (upper panel) and on the ``$-$'' (redshifted) side of the wormhole (lower panel). The redshift between ``$-$'' and ``$+$'' observers is $(1+z)^{-1}\approx0.942710$. The time is measured in units of the corresponding asymptotic observer. }\label{fig:profilepmb14}
\end{figure*}

\subsection{The brane-world model}

In \cite{Bronnikov:2019sbx} the second Randall-Sundrum brane-world model (RS2) \cite{Randall:1999vf} is considered implying that our four-dimensional world is a hypersurface supporting all matter fields and embedded in a ${\mathbb Z}_2$-symmetric five-dimensional spacetime (asymptotically AdS bulk), while the gravitational field propagates in the whole bulk. The gravitational field on the brane itself is described by the modified Einstein equations derived by Shiromizu, Maeda and Sasaki~\cite{Shiromizu:1999wj}
\begin{eqnarray}\label{EE4}
G_\mu^\nu = - \Lambda_4\delta_\mu^\nu -\kappa_4^2 T_\mu^\nu - \kappa_5^4 \Pi_\mu^\nu - E_\mu^\nu,
\end{eqnarray}
where $G_\mu^\nu = R_\mu^\nu - \frac{1}{2} \delta_\mu^\nu R$ is the 4D Einstein tensor, $\Lambda_4$ is the 4D cosmological constant expressed in terms of the 5D cosmological constant $\Lambda_5$ and the brane tension $\lambda$:
\begin{equation}\label{La4}
\Lambda_4 = \frac{1}{2}\kappa_5^2 \biggl(\Lambda_5 + \frac{1}{6} \kappa_5^2\lambda^2\biggr);
\end{equation}
$\kappa_4^2 = 8\pi G_N = \kappa_5^4 \lambda/(6\pi)$ is the 4D gravitational constant ($G_N$ is the Newtonian constant of gravity); $T_\mu^\nu$ is the stress-energy tensor of matter
located on the brane;
$\Pi_\mu^\nu$ is a tensor quadratic in $T_\mu^\nu$, obtained from the matching conditions for the 5D metric across the brane:
\begin{equation}\label{Pi_}
\Pi_\mu^\nu = \frac{1}{4} T_\mu^\alpha T_\alpha^\nu - \frac{1}{2} T T_\mu^\nu - \frac{1}{8} \delta_\mu^\nu \left( T_{\alpha\beta} T^{\alpha\beta} -\frac{1}{3} T^2\right),
\end{equation}
where $T = T^\alpha_\alpha$; lastly, $E_\mu^\nu$ is the so-called ``electric'' part of the 5D Weyl tensor projected onto the brane: in proper 5D coordinates, we have
$$E_{\mu\nu} = \delta_\mu^A \delta_\nu^C {}^{(5)} C_{ABCD} n^B n^D,$$
where the capital letters $A, B, \ldots$ are 5D indices, and $n^A$ is the unit normal vector to the brane.
By construction, $E_\mu^\nu$ is traceless, that is, $E_\mu^\mu = 0$ \cite{Shiromizu:1999wj}. The general class and a number of particular examples describing wormholes and black holes in the above brane-world scenario were obtained in \cite{Bronnikov:2002rn,Bronnikov:2003gx}.

The Bronnikov-Kim-1 (BK-1) metric can be written as
\begin{eqnarray}\nonumber
ds^2&=&-N^2(r)dt^2 + \frac{dr^2}{B^2(r)} + r^2(d\theta^2 + \sin^2\theta d\varphi^2),\\
N(r)&=&1-\frac{2 \mathcal{M}}{r}, \qquad B(r)=\sqrt{\left(1-\frac{r_0}{r}\right)\left(1-\frac{r_1}{r}\right)},\qquad
\label{f-x3}
\end{eqnarray}
where $r_1 = \mathcal{M} r_0/(r_0- \mathcal{M})$.

The only black-hole solution corresponds to the case $r_0 = r_1 = 2 \mathcal{M}$, which coincides with the extremal Reissner-Nordström metric.

Other values of $r_0$ lead either to wormholes (the throat is located at $r = r_0$ if $r_0 > 2 \mathcal{M}$ or at $r = r_1 > 2 \mathcal{M}$ in case $2 \mathcal{M} > r_0 > \mathcal{M}$), or to a naked singularity located at $r=2 \mathcal{M}$ (if $r_0 < \mathcal{M}$) (see more details in \cite{Bronnikov:2002rn}). Notice, that the above solution (\ref{f-x3}) is identical to (\ref{analyticsln}) up to the following redefinition of constants:
\begin{equation}
\mathcal{M}=\frac{M}{2}=\frac{Q_e^2r_0}{Q_e^2+r_0^2},\; \; r_1=\frac{Q_e^2}{r_0}.
\end{equation}

\section{Master wave equations}\label{sec:waveequations}

The general covariant equations for the scalar field $\Phi$, the electromagnetic field $A_\mu$ and the Dirac field $\Upsilon$ \cite{Brill:1957fx} are respectively written as:
\begin{subequations}\label{perturbeqs}
\begin{eqnarray}\label{KGg}
\frac{1}{\sqrt{-g}}\partial_\mu \left(\sqrt{-g}g^{\mu \nu}\partial_\nu\Phi\right)&=&0,\\
\label{EmagEq}
\frac{1}{\sqrt{-g}}\partial_{\mu} \left(F_{\rho\sigma}g^{\rho \nu}g^{\sigma \mu}\sqrt{-g}\right)&=&0\,,\\
\label{covdirac}
\gamma^{\mu} \left( \partial_{\mu} - \Gamma_{\mu} \right) \Upsilon&=&0,
\end{eqnarray}
\end{subequations}
where $F_{\mu\nu}=\partial_\mu A_\nu-\partial_\nu A_\mu$ 
and $\Gamma_{\mu}$ are spin connections. 
After separation of the variables Eqs.~(\ref{perturbeqs}) take the following Schrödinger-like form (see, for instance, \cite{Konoplya:2011qq,Kokkotas:1999bd})
\begin{equation}\label{wave-equation}
\dfrac{\partial^2 \Psi}{\partial t^2}-\dfrac{\partial^2 \Psi}{\partial r_*^2}=V(r)\Psi,
\end{equation}
where the ``tortoise coordinate'' $r_*$ is defined by the relation
\begin{equation}
dr_*=\frac{dr}{N(r)B(r)}.
\end{equation}

The effective potentials for the scalar ($s=0$) and electromagnetic ($s=1$) fields are
\begin{equation}\label{potentialScalar}
V(r)=N^2(r) \frac{\ell(\ell+1)}{r^2}+\left(1-s\right)\cdot\frac{1}{2r}\dfrac{d \left(N^2(r)B^2(r)\right)}{dr},
\end{equation}
where $\ell=0, 1, 2, \ldots$ are the multipole numbers.
For the Dirac field we have two isospectral potentials
\begin{equation}
V_{\pm}(r) = \frac{k}{r} \left(\frac{k N^2(r)}{r}\mp\frac{N^2(r)B(r)}{r}\pm N(r)B(r)\dfrac{d N(r)}{dr}\right),
\end{equation}
(where $k=1, 2, 3, \ldots$ are the multipole numbers, $k=\ell+1/2$) which can be transformed one into another by the Darboux transformation
\begin{equation}\label{psi}
\Psi_{+}=C (W+\dfrac{d}{dr_*}) \Psi_{-}, \quad W=\sqrt{N(r)B(r)}.
\end{equation}
Here $C$ is a constant. As $+$ and $-$ wave equations are isospectral, we can consider only one of the two effective potentials for the Dirac case.

\begin{table*}
\PRDstyle{\begin{tabular}{p{2cm}c@{\hspace{2cm}}c@{\extracolsep{\fill}}}}
\JCAPstyle{\begin{tabular}{p{2em}cc}}
\hline
$r_0$ & $\ell=0$ & $\ell=1$ \\
\hline
1   & $0.13314-0.09599i$  ($0.13384-0.09484i$) & $0.37760-0.08934i$ [$0.37765-0.08938i$] \\
1.1 & $0.12822-0.09481i$,\, echoes             & $0.37729-0.08845i$,\, echoes  \\
1.2 & $0.12691-0.08216i$,\, echoes             & $0.37487-0.08581i$,\, echoes  \\
1.3 & $0.12063-0.08325i$,\, $0.13995-0.01213i$ & $0.37136-0.07832i$,\, echoes  \\
1.4 & $0.11239-0.01474i$,\, $0.15335-0.01988i$ & $0.36950-0.06984i$,\, $0.35493-0.00487i$  \\
1.5 & $0.10930-0.00913i$,\, $0.16097-0.02709i$ & $0.36797-0.06917i$,\, $0.37670-0.01036i$  \\
\hline
\end{tabular}
\caption{Fundamental quasinormal mode of the scalar field in the background of the Kim-Bronnikov wormhole ($\mathcal{M}=1/2$) obtained by the time-domain method. The frequency in brackets is found by the higher order WKB method for the extreme Reissner-Nordström case.}\label{tabl:scalar}
\end{table*}

\begin{table*}
\PRDstyle{\begin{tabular}{p{2cm}c@{\hspace{2cm}}c@{\extracolsep{\fill}}}}
\JCAPstyle{\begin{tabular}{p{2em}cc}}
\hline
$r_0$ & $\ell=1$ & $\ell=2$ \\
\hline
1   & $0.33527-0.08435i$ ($0.33530-0.08439i$)  & $0.60144-0.08697i$ [$0.60144-0.08697i$] \\
1.1 & $0.33398 - 0.08207i$,\, echoes           & $0.59792-0.08353i$,\, echoes  \\
1.2 & $0.33257-0.07551i$,\, echoes             & $0.59725-0.08085i$,\, echoes  \\
1.3 & $0.32922-0.06451i$,\, echoes             & $0.59614-0.07754i$,\, echoes  \\
1.4 & $0.32110-0.04536i$,\, $0.33205-0.00745i$ & $0.59423-0.07374i$,\, echoes  \\
1.5 & $0.31726-0.04325i$,\, $0.35033-0.01396i$ & $0.59294-0.06635i$,\, $0.58678-0.00501i$  \\
\hline
\end{tabular}
\caption{Fundamental quasinormal mode of the electromagnetic field in the background of the Kim-Bronnikov wormhole ($\mathcal{M}=1/2$) obtained by the time-domain method. The frequency in brackets is found by the higher order WKB method for the extreme Reissner-Nordström case.}\label{tabl:em}
\end{table*}

\begin{table*}
\PRDstyle{\begin{tabular}{p{2cm}c@{\hspace{2cm}}c@{\extracolsep{\fill}}}}
\JCAPstyle{\begin{tabular}{p{2em}cc}}
\hline
$r_0$ & $k=1$ & $k=2$ \\
\hline
1   & $0.23825-0.08754i$ ($0.23809-0.08757i$)  & $0.49411-0.08824i$ [$0.49411-0.08824i$] \\
1.1 & $0.23798-0.08641i$,\, echoes             & $0.49306-0.08686i$,\, echoes  \\
1.2 & $0.23465-0.07191i$,\, echoes             & $0.49029-0.08106i$,\, echoes  \\
1.3 & $0.23370-0.06236i$,\, $0.23865-0.01073i$ & $0.46923-0.06411i$,\, echoes  \\
1.4 & $0.23122-0.06135i$,\, $0.25699-0.02088i$ & $0.46420-0.05861i$,\, $0.46801-0.00452i$  \\
1.5 & $0.22898-0.05279i$,\, $0.26592-0.03082i$ & $0.46102-0.05626i$,\, $0.49381-0.01104i$  \\
\hline
\end{tabular}
\caption{Fundamental quasinormal mode of the Dirac field in the background of the Kim-Bronnikov wormhole ($\mathcal{M}=1/2$ obtained by the time-domain method. The frequency in brackets is found by the higher order WKB method for the extreme Reissner-Nordström case.}\label{tabl:Dirac}
\end{table*}

\section{Quasinormal modes and echoes}\label{sec:QNMs}

Quasinormal modes $\omega_{n}$ correspond to solutions of the master wave equation (\ref{wave-equation}) with the requirement of the purely outgoing waves at both infinities \cite{Konoplya:2005et}:
\begin{equation}
\Psi \propto e^{-\imo \omega t \pm \imo \omega r_*}, \quad r_* \to \pm \infty.
\end{equation}

For finding of the low-laying quasinormal modes we will mainly use the time-domain integration method. One can integrate the wavelike equation (\ref{wave-equation}) in terms of the light-cone variables $u=t-r_*$ and $v=t+r_*$. We apply the discretization scheme proposed in \cite{Gundlach:1993tp},
\begin{eqnarray}\label{Discretization}
\Psi\left(N\right)&=&\Psi\left(W\right)+\Psi\left(E\right)-\Psi\left(S\right)\PRDstyle{\\\nonumber&&}
-\Delta^2V\left(S\right)\frac{\Psi\left(W\right)+\Psi\left(E\right)}{4}+{\cal O}\left(\Delta^4\right)\,,
\end{eqnarray}
where we used the following notation for the points:
$N\equiv\left(u+\Delta,v+\Delta\right)$, $W\equiv\left(u+\Delta,v\right)$, $E\equiv\left(u,v+\Delta\right)$, and $S\equiv\left(u,v\right)$. The Gaussian initial data are imposed on the two null surfaces, $u=u_0$ and $v=v_0$. We extract the dominant quasinormal frequencies from the time-domain profiles using the Prony method \cite{Prony}.

For the extreme black-hole state the effective potential has only one peak. Therefore, we will also use the WKB method of Will and Schutz \cite{Schutz:1985zz}, which was extended to higher orders in \cite{Iyer:1986np,Konoplya:2003ii,Matyjasek:2017psv} and made even more accurate by the usage of the Padé approximants in \cite{Matyjasek:2017psv,Hatsuda:2019eoj}.
The higher-order WKB formula has the form \cite{Konoplya:2019hlu},
\begin{eqnarray}
 \omega^2&=&V_0+A_2(\K^2)+A_4(\K^2)+A_6(\K^2)+\ldots \\\nonumber
&-& \imo \K\sqrt{-2V_2}\left(1+A_3(\K^2)+A_5(\K^2)+A_7(\K^2)+\ldots\right),
\end{eqnarray}
where $\K$ takes half-integer values. The corrections $A_k(\K^2)$ of the order $k$ to the eikonal formula are polynomials of $\K^2$ with rational coefficients and depend on the values of higher derivatives of the potential $V(r)$ in its maximum. In order to increase accuracy of the WKB formula, we follow Matyjasek and Opala \cite{Matyjasek:2017psv} and use the Padé approximants. Here we will use the sixth order WKB method with $\tilde{m} =5$, where $\tilde{m}$ is defined in \cite{Matyjasek:2017psv,Konoplya:2019hlu}, because this choice provides the best accuracy for the black-hole limit.

As both methods are extensively discussed in the literature (see, for example, reviews \cite{Konoplya:2019hlu,Konoplya:2011qq}), we will not describe them in this paper, but will show that both methods are in a good agreement in the common parametric range of applicability.

Here we will distinguish the analytical solution given by formulas (\ref{analyticsln}) and (\ref{f-x3}), which appear in both theories under consideration and the numerical solutions which appear in the Einstein-Maxwell-Dirac theory with massive spinors. The Einstein-Maxwell-Dirac theory solutions have nonzero background Dirac and Maxwell fields, which, therefore, cannot be considered as test fields in this approach. The analytical solution can also be interpreted in the brane-world model, where no coupled Maxwell and Dirac fields are added. Therefore, we can consider test Maxwell and Dirac fields in the background (\ref{f-x3}) having in mind the brane-world scenario. In addition, the analytical solution in the context of Einstein-Maxwell-Dirac, can be considered as a good approximation to the wormhole configuration supported by neutral massive fermions, when the fermion mass is sufficiently small, for example, for neutrinos.

\begin{figure}
\resizebox{\linewidth}{!}{\includegraphics{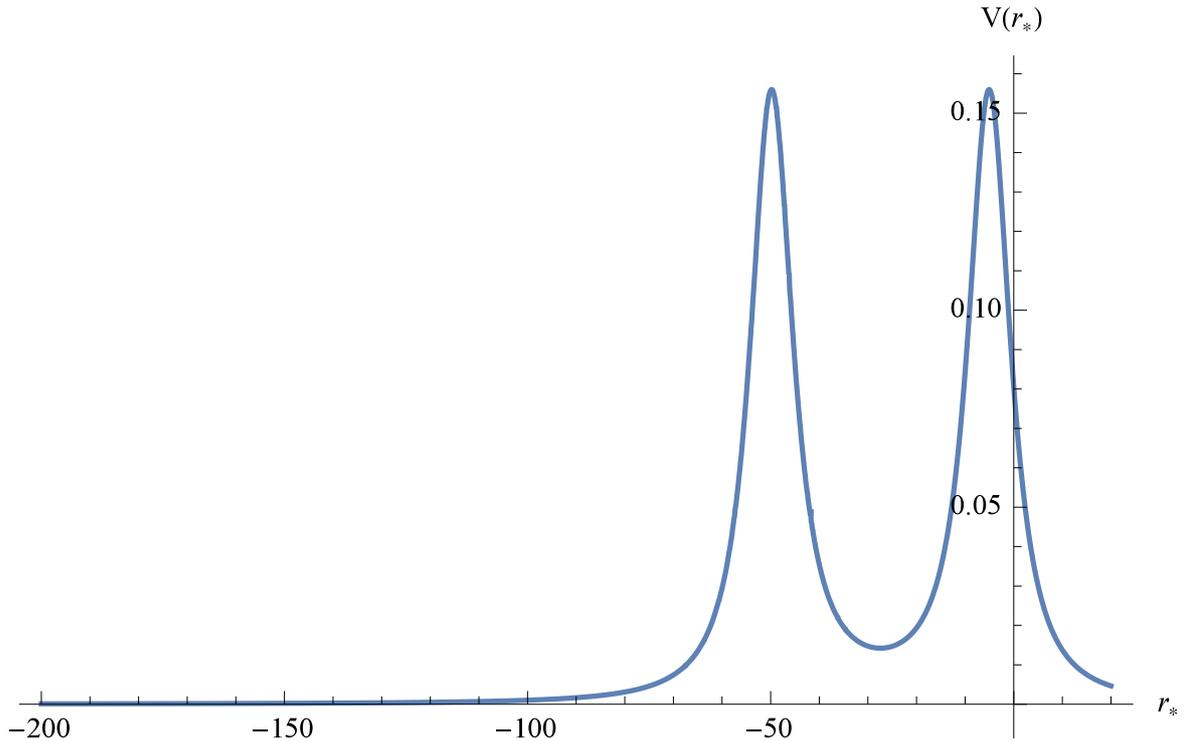}}
\caption{Effective potential for the scalar field ($\ell=1$) in the background of the Kim-Bronnikov wormhole $r_0=1.1$ ($\mathcal{M}=1/2$).}\label{fig:PotentialScalar1p11}
\end{figure}

From tables \ref{tabl:scalar}, \ref{tabl:em}, and \ref{tabl:Dirac} we can see that once the throat radius $r_{0}$ approaches $2 \mathcal{M}$ from above, the fundamental quasinormal mode approaches the extreme Reissner-Nordström value for all types of perturbations (scalar, electromagnetic, Dirac) under consideration. In this extreme limit, the quasinormal modes are computed also with the WKB method and the agreement with the time-domain integration is within the numerical accuracy of the quasinormal frequencies extracted from time-domain profiles. Both the real oscillation frequency of the dominant mode and the damping rate increase as the wormhole configuration approaches the extreme (black hole) state $2 \mathcal{M}$. Near the wormhole/black hole transition the echoes \cite{Cardoso:2016rao,Annulli:2021dkw} appear as the result of sub-scattering from the second peak of the effective potential (see fig.~\ref{fig:PotentialScalar1p11}). These echoes are more distinctive than those from the possible environmental matter at a distance from the compact object \cite{Konoplya:2018yrp} and thereby give a clear indication of the wormhole geometry. Sufficiently far from the transition point, the second mode can be seen instead of the peak, which, apparently, appears after a very long time of quasinormal decay. Notice, that the data for the $\ell=1$ electromagnetic perturbations presented earlier by one of us for the Kim-Bronnikov wormhole (table~II in \cite{Bronnikov:2019sbx}) contained a numerical mistake and here we correct this.

The quasinormal spectrum of the wormhole supported by massive fermions has a number of peculiar properties, because the wormhole is asymmetric relatively the throat and different positions of the observer and location of the initial gaussian wave package must be considered. As the wormhole solution is known only numerically in this case, we had to integrate the wave equation with very high accuracy and small integration step in order not to accumulate the numerical error at late times.

Time-domain profile for the spherically symmetric ($\ell=0$) test scalar field in the background of the asymmetric wormhole ($b_i=0.14$) has been constructed in fig. \ref{fig:profilepmb14}. On the upper panel the maximum of the initial wave package of perturbation and the distant observer are located on the ``$+$'' (blueshifted) side of the wormhole.
The first ringing corresponds to the near extreme Reissner-Nordström black hole with the dominant frequency $\omega r_0\approx 0.1338 - 0.0961\imo$ (cf.~Table~\ref{tabl:scalar}). The first echo signal contains two dominant quasinormal modes, $\omega r_0\approx0.124 - 0.078\imo$ and $\omega r_0\approx0.147 - 0.088\imo$. The next echoes are governed by the modes with smaller real and imaginary parts: the modes for the second echoes are $\omega r_0\approx0.122 - 0.070\imo$ and $\omega r_0\approx0.147 - 0.075\imo$.
A similar picture is obtained on the ``$-$'' (redshifted) side of the wormhole (shown on the lower panel): first we observe the Reissner-Nordström ringing with the dominant frequency $\omega r_0\approx 0.1338 - 0.0961\imo$, while the echo signal contains two dominant modes, $\omega r_0\approx0.131-0.081\imo$ and $\omega r_0\approx0.153-0.093\imo$ (first echo) and $\omega r_0\approx0.126-0.072\imo$ and $\omega r_0\approx0.153-0.076\imo$ (second echo).

We measure the time variable in the units of the corresponding asymptotic observer. We notice that, when we take into account the redshift, the characteristic time until the echo appears is the same and the echoes on both sides of the wormhole are governed by the same quasinormal modes. However, we did not obtain the accurate values for the second mode of the second echo. We can see that the echo is the true footprint of the wormhole, while the dominant quasinormal mode is very close to that for the extreme Reissner-Nordström black hole.

\begin{table}
\JCAPstyle{
\begin{tabular}{ccccccccc}
\hline
$b_i$ & $g_i$ & $f_i$ & $(1+z)^{-1}$ & $M_+/r_0$ & $M_-/r_0$ & $R_{\rm sh+}/r_0$ & $R_{\rm sh-}/r_0$ \\
\hline
 0.02 & 0.0021907 & -0.0027540 & 0.945754 & 0.999953 & 0.999948 & 3.99981 & 3.99979 \\
 0.03 & 0.0032861 & -0.0041313 & 0.945685 & 0.999894 & 0.999883 & 3.99957 & 3.99953 \\
 0.04 & 0.0043819 & -0.0055093 & 0.945577 & 0.999811 & 0.999792 & 3.99924 & 3.99917 \\
 0.05 & 0.0054780 & -0.0068882 & 0.945436 & 0.999705 & 0.999676 & 3.99882 & 3.99870 \\
 0.06 & 0.0065746 & -0.0082681 & 0.945261 & 0.999575 & 0.999533 & 3.99830 & 3.99813 \\
 0.07 & 0.0076718 & -0.0096493 & 0.945053 & 0.999421 & 0.999364 & 3.99768 & 3.99745 \\
 0.08 & 0.0087696 & -0.0110319 & 0.944816 & 0.999244 & 0.999169 & 3.99697 & 3.99667 \\
 0.09 & 0.0098682 & -0.0124162 & 0.944544 & 0.999043 & 0.998948 & 3.99616 & 3.99578 \\
 0.10 & 0.0109675 & -0.0138024 & 0.944236 & 0.998819 & 0.998701 & 3.99526 & 3.99479 \\
 0.11 & 0.0120678 & -0.0151907 & 0.943892 & 0.998571 & 0.998429 & 3.99426 & 3.99369 \\
 0.12 & 0.0131692 & -0.0165813 & 0.943516 & 0.998300 & 0.998129 & 3.99317 & 3.99248 \\
 0.13 & 0.0142715 & -0.0179746 & 0.943093 & 0.998005 & 0.997804 & 3.99197 & 3.99117 \\
 0.14 & 0.0153756 & -0.0193696 & 0.942710 & 0.997686 & 0.997453 & 3.99068 & 3.98974 \\
 0.15 & 0.0164801 & -0.0207698 & 0.942144 & 0.997345 & 0.997075 & 3.98930 & 3.98821 \\
 0.20 & 0.0220280 & -0.0278195 & 0.939077 & 0.995284 & 0.994792 & 3.98087 & 3.97886 \\
 0.25 & 0.0276233 & -0.0349855 & 0.934792 & 0.992641 & 0.991843 & 3.96989 & 3.96659 \\
 0.30 & 0.0332790 & -0.0423078 & 0.928984 & 0.989420 & 0.988211 & 3.95624 & 3.95113 \\
 0.35 & 0.0390074 & -0.0498370 & 0.921134 & 0.985632 & 0.983863 & 3.93974 & 3.93204 \\
 0.40 & 0.0448190 & -0.0576415 & 0.910381 & 0.981288 & 0.978740 & 3.92013 & 3.90859 \\
 0.45 & 0.0507260 & -0.0657993 & 0.895754 & 0.976401 & 0.972724 & 3.89704 & 3.87947 \\
 0.50 & 0.0566841 & -0.0745429 & 0.872497 & 0.971041 & 0.965526 & 3.87022 & 3.84183 \\
 \hline
 0.02 & 0.0016663 & 0.00116784 & 0.994789 & 0.999886 & 0.999881 & 3.99954 & 3.99952 \\
 0.03 & 0.0024979 & 0.00175177 & 0.995020 & 0.999743 & 0.999732 & 3.99897 & 3.99893 \\
 0.04 & 0.0033319 & 0.00233509 & 0.995076 & 0.999542 & 0.999523 & 3.99817 & 3.99809 \\
 0.05 & 0.0041673 & 0.00291792 & 0.995146 & 0.999284 & 0.999255 & 3.99713 & 3.99702 \\
 0.06 & 0.0050044 & 0.00350009 & 0.995230 & 0.998969 & 0.998926 & 3.99587 & 3.99570 \\
 0.07 & 0.0058436 & 0.00408146 & 0.995328 & 0.998595 & 0.998537 & 3.99437 & 3.99414 \\
 0.08 & 0.0066851 & 0.00466188 & 0.995444 & 0.998163 & 0.998087 & 3.99263 & 3.99233 \\
 0.09 & 0.0075229 & 0.00524883 & 0.995455 & 0.997673 & 0.997578 & 3.99066 & 3.99027 \\
 0.10 & 0.0083434 & 0.00585859 & 0.995159 & 0.997123 & 0.997040 & 3.98845 & 3.98798 \\
\hline
\end{tabular}
}
\caption{Shadow radius $R_{sh\pm}$ for two sides of the asymmetric wormholes supported by the massive charged fermions ($qr_0=0.03$ and $\mu r_0=0.2$).}\label{tabl:shadowradius}
\end{table}

\begin{figure}
\resizebox{\linewidth}{!}{\includegraphics{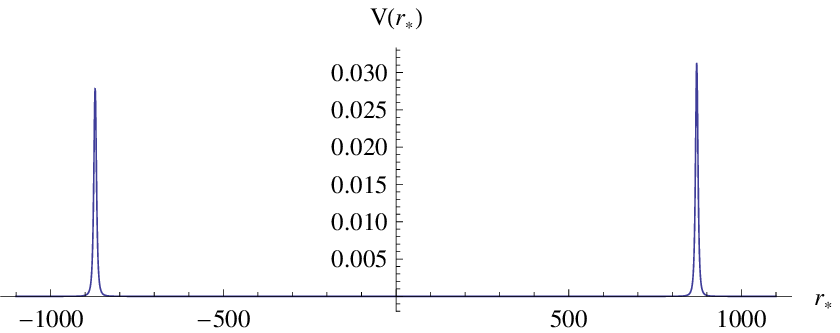}}
\resizebox{\linewidth}{!}{\includegraphics{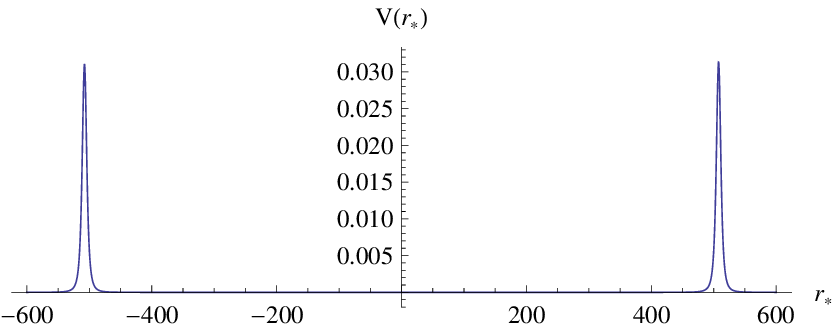}}
\caption{Effective potential for the $\ell=0$ scalar field perturbations in the background of the asymmetric wormhole supported by the massive charged fermions ($qr_0=0.03$ and $\mu r_0=0.2$) for $b_i=0.08$:
the fermion configurations correspond to $f_i=-0.0110319$ and $g_i=0.0087696$ (top panel) and to $f_i=0.0046619$ and $g_i=0.0066851$ (bottom panel). The time is measured according to the ``$+$'' (blueshifted) observer and the dimensional quantities are given in units of the wormhole's throat radius ($r_0=1$). }\label{fig:awpot}
\end{figure}

Notice that the two families of wormhole solutions were numerically obtained and studied in \cite{Konoplya:2021hsm}. From fig. \ref{fig:awpot} one can see that while the profiles of the peaks of effective potentials are similar for both families of solutions, the separation between the two peaks is smaller for the second family of solutions. Therefore, the time needed for the second reflection from the peak is shorter and the echoes appear earlier for the second family of solutions. The quasinormal frequencies during the main ringdown phase and during echoes are very close for both families.

\section{Shadows}\label{sec:shadows}

The shadow radius of an arbitrary static, spherically symmetric compact object \cite{Synge:1966okc}, including wormhole solutions can be computed according to the procedure described in \cite{Bronnikov:2021liv}. The shadow is closely related to the location of its photon sphere $r_{\rm ph}$. It is convenient to introduce the function
\begin{equation}
w(r) \equiv \frac{r}{N(r)}\,,
\end{equation}
so that $r_{\rm ph}$ is the minimum of $w(r)$ and determined as the solution to the equation
\begin{equation}
\frac{d w}{dr}=0\,.
\label{uph_condition}
\end{equation}
Next, the angular radius of the shadow, as seen by a distant static observer located at
$r_{\rm O}$, can be found by the following relation
\begin{equation}
\sin{a_{\rm sh}} = \frac{w(r_{\rm ph})}{w(r_{\rm O})}\,.
\end{equation}
Under the assumption that the observer is located far from the compact
object, i.e., $r_{ \rm O} \gg r_0$, where $r_0$ is a characteristic length scale that can be
identified with the radius of the wormhole throat, we have:
\begin{equation}
N(r_{\rm O}) \simeq 1.
\end{equation}
Then one finds that the radius of the shadow is given by
\begin{equation}
R_{\rm sh} \simeq r_{\rm O} \sin{a_{\rm sh}} \simeq w(r_{\rm ph})= \frac{r_{\rm ph}}{N(r_{\rm ph})}\,.
\label{Rsh}
\end{equation}
Notice that the shadow radius of an asymmetric wormhole will be different when measured by distant observers on the left and right from the throat.

Here we complement previous study of the shadow radius of the analytical solution which has been performed in~\cite{Bronnikov:2021liv} by extending the analysis to the numerical solution supported by massive fermions~\cite{Konoplya:2021hsm}.
From table~\ref{tabl:shadowradius} we can see that the shadow radius is slightly larger when measured by a distant observer on the wormhole's side corresponding to a larger asymptotic mass. This occurs for both families of solutions. When the parameter $b_i$ increases, the radius of the shadow is slightly decreasing, staying near the extreme Reissner-Nordström value.

\section{Conclusions}\label{sec:conclusions}

Here, for the first time, quasinormal modes, echoes and radius of the shadow were studied for a class of wormholes constructed in General Relativity without using exotic matter~\cite{Blazquez-Salcedo:2020czn,Konoplya:2021hsm}. The wormhole configuration is supported by fermion and electromagnetic fields. When the fermion is neutral and massless, the analytical solution exists~\cite{Blazquez-Salcedo:2020czn} and it coincides (up to the redefinition of constants) with the metric obtained in the brane-world model~\cite{Bronnikov:2002rn}. Having in mind this application we also studied quasinormal frequencies of test fields of various spins in the background of this analytical solution. The quasinormal spectrum of the analytical solution strongly depends upon the wormhole parameters. The shadows and quasinormal ringing are studied also for the smooth asymmetric wormholes supported by massive fermions. Both the quasinormal modes and shadows are close to those for the extreme Reissner-Nordström black hole in the considered range of parameters. Nevertheless, the echoes give a distinctive footprint of the wormhole geometry in all cases. The asymptotic observers on two sides of the asymmetric wormholes observe that the echo is governed by the same quasinormal frequencies if they take into account their relative redshift.

The important question which was outside our consideration is the spectrum and stability of gravitational perturbations, which would allow us to judge on stability of the wormhole itself. In order to study the linear stability of the wormhole, one has to perform perturbation of all fields in the system and reduce the perturbation equations to a set of differential master wavelike equations.

\begin{acknowledgments}
We acknowledge support of the grant 19-03950S of Czech Science Foundation (GAČR). We would like to thank Robert Wald and Robert Weinbaum for useful discussions.
\end{acknowledgments}

\end{document}